\documentclass[usenatbib]{mn2e}
\usepackage{natbib}
\usepackage{amsmath}
\usepackage{amssymb}
\usepackage{graphicx}
\usepackage{color}
\usepackage[colorlinks=true,linkcolor=darkblue,citecolor=darkblue,urlcolor=darkblue]{hyperref}
\usepackage{epstopdf}

\usepackage{verbatim}

\usepackage{tikz}   
\usepackage{tikz-3dplot} 

\newcommand{\AxisRotator}[1][rotate=0]{%
    \tikz [x=0.15cm,y=0.30cm,line width=.2ex,-stealth,#1] \draw (0,0) arc (-150:150:1 and 1);%
}

\definecolor{darkblue}{rgb}{0.05,0.0,0.5}

\begin{document}
\title[The spin of the black hole in galactic centre]{Probing the spin of the central black hole in the galactic centre with secondary images}
\author[Jonas Helboe J{\o}rgensen, Ole Eggers Bj{\ae}lde and Steen Hannestad]{Jonas Helboe J{\o}rgensen$^1$\thanks{helboe@phys.au.dk}, Ole Eggers Bj{\ae}lde$^1$\thanks{oeb@phys.au.dk} and Steen Hannestad$^1$\thanks{sth@phys.au.dk}\\
$^1$Department of Physics and Astronomy, Aarhus University \\ Ny Munkegade 120, DK-8000 Aarhus C, Denmark}

\maketitle
\begin{abstract}
This paper explores the possibility of determining the spin of the supermassive black hole (SMBH) in Sgr A*, by using secondary images of stars orbiting the SMBH. The photons propagate close to the SMBH and their trajectories probe the space time in a region where the spin of the SMBH is important. We find the appearance of spikes in the secondary image, which depends on the angular momentum and spin axis of the SMBH and study the specific case of the star S2 in detail. 
The spikes has a magnitude of $\sim 29$ in the K-band and the required angular resolution is of order 15-20 $\mu$as. The combination of these two requirements poses an extreme observational challenge, but might be possible with interferometric observations in the sub-mm regime. The next possible time frame for observing this effect on the star S2 is in the late 2017 and then it repeats with the period of the star.
\end{abstract}

\section{Introduction}
The possible rotation of supermassive black holes (SMBHs) have been a subject of much interest lately (\cite{Reynolds:2013rva, Angelil:2014yea, Jiang:2015dla, Christian:2015iia, Brink:2015roa}). The SMBH in Sgr A* in the center of the Milky Way galaxy is no exception but observational difficulties within the galaxy complicates the process. In this paper we suggest a new approach to observe the possible rotation of the SMBH in the galactic centre.

Black holes in general are described by the Kerr metric which describes a space time around a rotating spherically symmetric mass-distribution. In the Kerr metric, a black hole is fully described by its mass $M$ and angular momentum $a$. The masses of black holes are often straightforwardly determined from the dynamics of stars and other objects in the vicinity of the black hole - see e.g. \cite{Ghez:2008ms, Gillessen:2008qv, Zuo:2014fea}, the rotation of black holes, however, is harder to determine since it only affects the metric close to the black hole event horizon (\cite{Melia:2001fp}). Previous works have determined the rotation of SMBHs  (\cite{Broderick:2008sp,Reynolds:2013rva}) by observing the perturbing effect of rotation on accretion flow or accretion discs around black holes (see also (\cite{Li:2013jra}) for an example of a rotating stellar-sized black hole) and have indicated that SMBHs with masses in the region of $10^6\, M_\odot$ have typical angular momenta near their maximally allowed values by the Kerr metric (see however \cite{Broderick:2010kx} for a different conclusion).

In the Milky Way there is overwhelming evidence that a SMBH is associated with the object Sgr A* (see e.g.\ (\cite{Maoz:1997yd}) for constrains on alternatives to a black hole in the center of the galaxy). While the mass of the SMBH has been determined accurately by observing orbits of stars around the SMBH (\cite{Gillessen:2008qv, Meyer:2012hn, 2014IAUS..303..242B}), the rotation has not yet been measured precisely. Determining the rotation of the SMBH would give clues of its origin (\cite{1996MNRAS.283..854M,Volonteri:2004cf}). 

In this paper we investigate the use of secondary images, defined as images where light has traveled behind the black hole, to probe the region of the spacetime around the SMBH in the galactic centre. We focus on the secondary image of the star S2 and determine its dependence on the angular momentum and spin direction of the SMBH. The trajectory of photons associated with this image propagate very near to the SMBH and will thus be affected by the metric close to the black hole. Present day technology is insufficient to observe the secondary image of S2, however, future observational projects such as The Event Horizon Telescope may have a sufficiently large aperture to observe the secondary image. 

\section{Simulation}
The objective is to find the temporal evolution of the location and magnitude of secondary images of stars in the S-star cluster orbiting the SMBH in the center of our galaxy. The parameters for stellar orbits in the S cluster as well as mass of and distance to the central SMBH will be taken from (\cite{Gillessen:2008qv})
\begin{align*}
M_{\rm SMBH}&=(4.31\pm0.06|_{stat}\pm0.36|_{R0} )\cdot 10^6 M_{\odot},\\
d_{\rm SMBH}&=(8.33\pm0.35 )kpc,
\end{align*}
where the largest uncertainty on the mass comes from the uncertainty on the distance to the SMBH.

In the simulation, photons are propagated from the observer through the lens (SMBH) to a grid where the data are collected. This grid is designed such that the trajectory of S2 lays on the grid.

To describe a rotating black hole we use the Kerr metric which is given as,
\begin{align}
ds^2=&-\left(1- \frac{r_{s}r}{\Sigma} \right) \cdot c^2 dt^2- \frac{2r_{s}a \,r\sin^2\chi}{\Sigma} c dt d\xi + \frac{\Sigma}{\Delta} dr^2 \notag \\
&  + \Sigma d\xi ^2 +\left(r^2+a^2+\frac{r_s a^2 r \sin^2 \chi}{\Sigma}\right) \sin^2\chi d\xi^2. 
\end{align}
In this notation the following has been introduced,
\begin{align*}
& \Sigma=r^2+a^2  \cos^2\chi \\
& \Delta = r^2 - r_{s} r +a^2 \\
& r_{s} =\frac{2 G M}{c^2}.
\end{align*}
In the metric we have three coordinates $r, \, \chi $ and $\xi$. The metric is thus fully described by the mass $M$ of the black hole and the angular momentum per unit mass $a=\frac{J}{M}$. In this notation $r_s$ is the Schwarzschild radius of the corresponding non-rotating black hole.
The cosmic censorship hypothesis sets an upper limit to the rotation of a black hole to $a<\frac{r_s}{2}$  (\cite{Penrose:1969pc, Wald:1984}). The coordinate system is constructed such that the angular momentum vector is pointing towards $\chi=0$.

\begin{figure}
\centering
\tdplotsetmaincoords{60}{110}

\pgfmathsetmacro{\rvec}{.8}
\pgfmathsetmacro{\thetavec}{30}
\pgfmathsetmacro{\phivec}{50}

\begin{tikzpicture}[scale=5,tdplot_main_coords]
\coordinate (O) at (0,0,0);

\tdplotsetcoord{P}{\rvec}{\thetavec}{\phivec}

\draw[thick,->] (0,0,0) -- (1,0,0) node[anchor=north east]{$x$};
\draw[thick,->] (0,0,0) -- (0,1,0) node[anchor=north west]{$y$};
\draw[thick,->] (0,0,0) -- (0,0,1) node[anchor=south]{$z$};
\node[draw=none,shape=circle,fill, inner sep=5pt] (d1) at (0,0,0) {};  
\node at (-0.1,-0.3,0) {SMBH};
\node[draw=none,shape=circle,fill, inner sep=2pt] (d2) at (0.8,0,0){};  
\node at (0.8,-0.2,0) {Sun};

\draw[-stealth,color=red] (O) -- (P) node [midway] {\AxisRotator[rotate=60]};

\draw[dashed, color=red] (O) -- (Pxy);
\draw[dashed, color=red] (P) -- (Pxy);
%
\tdplotdrawarc{(O)}{0.2}{0}{\phivec}{anchor=north}{$\phi_A$}

\tdplotsetthetaplanecoords{\phivec}

\tdplotdrawarc[tdplot_rotated_coords]{(0,0,0)}{0.5}{0}{\thetavec}{anchor=south west}{$\theta_A$}

\draw[dashed,tdplot_rotated_coords] (\rvec,0,0) arc (0:90:\rvec);
\draw[dashed] (\rvec,0,0) arc (0:90:\rvec);

   \tdplotgetpolarcoords{0}{1}{0}
    \tdplotsetthetaplanecoords{\tdplotresphi}
     \pgfmathsetmacro{\radius}{0.1}
    \tdplotdrawarc[tdplot_rotated_coords, color=blue]{(0,0,-0.8)}{\radius}{0}%
    {360}{}{}
        \pgfmathsetmacro{\radius}{0.1}
    \tdplotdrawarc[tdplot_rotated_coords, color=blue]{(0,0,0)}{\radius}{0}%
    {360}{}{} 
    
 \draw[tdplot_rotated_coords, color=blue] 
     ({(sin(150))*0.1},
      {(cos(150))*0.1},0.0)
       -- 
     ({(sin(150))*\radius},
      {(cos(150))*\radius},
       -0.8) 
     node[font = \small, pos = 1.1] {};
      \draw[tdplot_rotated_coords, color=blue] 
     ({(sin(-30))*0.1},
      {(cos(-30))*0.1},0.0)
       -- 
     ({(sin(-30))*\radius},
      {(cos(-30))*\radius},
       -0.8) 
     node[font = \small, pos = 1.2] {Grid};

\end{tikzpicture}
\caption{The direction of the angular momentum vector of the SMBH is determined by the two angles $\phi_A$ and $\theta_A$. In this figure the galactic plane is the xy-plane and the solar system is fixed on the x axis at $(r_0,0,0)$. The cylindrical grid is shown in blue and lies parallel to the axis between the SMBH and earth. The cross section of the grid is made such that the stars orbit lies on surface of the cylinder.} \label{fig:skema}
\end{figure}
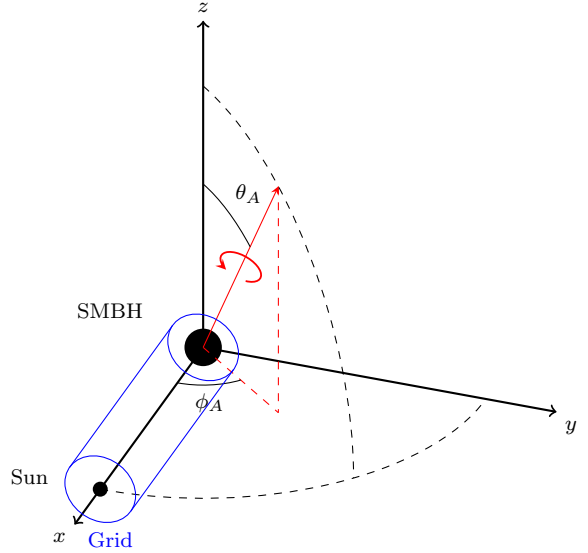

Photons are tracked back from the Sun to the SMBH and stopped on an elliptic cylindrical grid. The grid is placed such that the Sun and SMBH are on the central axis inside the elliptical cylinder and the stellar orbit is on the surface of the elliptical cylinder, shown on Fig.~\ref{fig:skema}. 
We propagate the photons backwards through the metric by use of the geodesic equation,
\begin{equation}
\frac{d^2x^\alpha}{d \lambda^2}=-\Gamma^{\alpha}_{\beta \gamma} \frac{dx^\beta}{d \lambda}\frac{dx^\gamma}{d \lambda},
\end{equation}
where the Cristoffel symbols $\Gamma^{\alpha}_{\beta \gamma}$ are given by the metric as,
\begin{equation}
\Gamma^{\alpha}_{\beta \gamma} =\frac{1}{2}g^{\alpha \delta} \left(\frac{\partial g_{\delta \beta}}{\partial x^{\gamma}} + \frac{\partial g_{\delta \gamma}}{\partial x^{\beta}} -\frac{\partial g_{\beta \gamma}}{\partial x^{\delta}}  \right).
\end{equation}
The Kerr metric and the non-zero Cristoffel symbols in the Kerr metric can be found in e.g. \cite{Muller:2009bw}.

As a technical detail, the angular momentum parameter $a$ has to change sign in the code, since the photons are being propagated backwards in time. 

We have two coordinate systems the one belonging to the metric where the angular momentum is at $\chi=0$, and one belonging to the galactic plane where the galactic plane is at $\theta=\pi/2$, see Fig.\ref{fig:skema}. In Fig.~\ref{fig:skema} the direction of the angular momentum is defined by the angles $\phi_A$ and $\theta_A$.
In the simulation we always start from the sun ($\phi=0$ and $\theta=\pi/2$). This can be transformed into the coordinate system of $\xi$ and $\chi$ by using the angles $\theta_A$ and $\phi_A$. 

\subsection*{Magnification}
By propagating one photon the position of the secondary image can be determined. To find the magnification of the secondary image we propagate three photons.
The magnification is determined as the relation between the solid angle of the image with and without the gravitational lensing, so $\mu=\frac{\Omega}{\Omega_0}$. To find this we propagate three photons back through the metric. The first photon is propagated with an initial direction of  $(b_x, b_y)$, where $(0,0)$ is directly at the black hole. The other two photons have initial directions of $(b_x+\delta b, b_y)$ and $(b_x, b_y+ \delta b)$, respectively, where $\delta b\ll b_i$. After the first photon is stopped at the cylindrical grid it defines a new plane perpendicular to the vector between the point and the black hole. The remaining two photons are then stopped on that plane and we define the positions of the two photons as $(d_{1x},d_{1y})$ and $(d_{2x},d_{2y})$, in a coordinate system with the first photon at $(0,0)$.
We can then find the magnification by finding the eigenvalues of the matrix.
\begin{equation}
T= \frac{1}{\delta b}\left( \begin{array}{cc} d_{1x} & d_{1y} \\ d_{2x} & d_{2y} \end{array} \right)
\end{equation}
The magnification of the image is given by the eigenvalues as
\begin{equation}
\mu= \frac{1}{\lambda_1 \lambda_2},
\end{equation}
where $\lambda_i$ are the eigenvalues of $T$. The magnitude of the secondary image is then found by using the magnification and the magnitude of the star.

\section{Results}
\begin{figure*}
\includegraphics[width=0.95\linewidth]{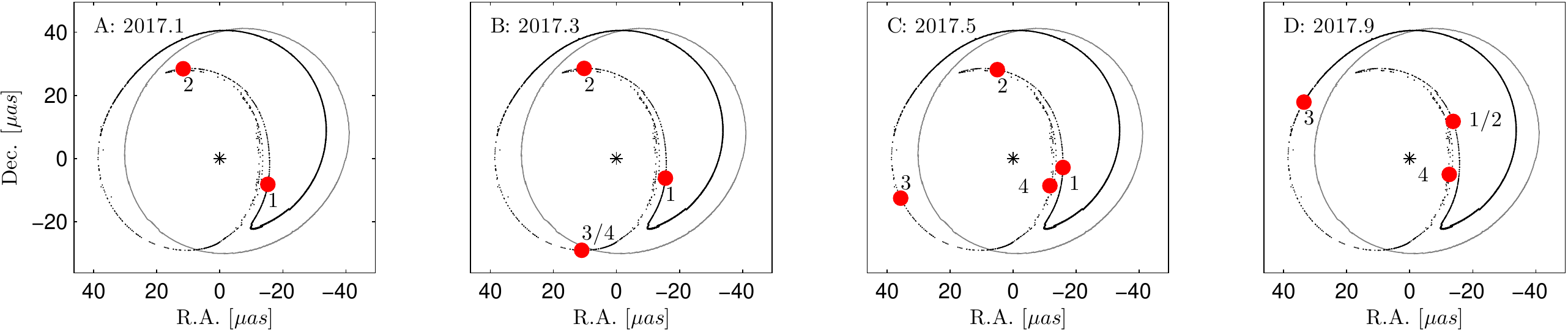}
\caption{Four identical images displaying the position of the secondary and tertiary images in terms of declination and right ascension for a black hole with $a=0$ and $a=0.99$, where the angular momentum vector is perpendicular to the galactic plane. The solid grey line indicates the orbit of the secondary image for $a=0$ and the black dots represent positions for the secondary and tertiary image for a black hole with $a=0.99$. The black dots form a line at the regions where the secondary image is at its peak.
The position of the secondary and tertiary images are overplotted with large red dots for four different epochs in 2017. In image $A$ the epoch is 2017.1 and the secondary image sits at (1) and the tertiary image at (2). In $B$ the epoch is 2017.3 and the secondary image sits at (1), the tertiary image at (2) and an image at (3/4) appears with high luminosity, corresponding to the spike (3/4) in Fig~\ref{fig:magnitude}. In $C$ the epoch is 2017.5 and the secondary image sits at (1), the tertiary image at (2) and the image at (3/4) has split into two images at (3) and (4). Finally in $D$ the epoch is 2017.9 and the two images (1) and (2) have merged into a single image with high luminosity, corresponding to the spike (1/2) in Fig~\ref{fig:magnitude}, which disappears immediately thereafter. Image (3) is now the new secondary image, which subsequently follows in solid line and becomes image (1) in image A. Image (4) is the new tertiary image and ends up as image (2) in image A. The magnitude of the images can be seen in Fig~\ref{fig:magnitude} at the different times.}
\label{fig:movement}
\end{figure*}
\begin{figure}
\includegraphics[width=1\linewidth]{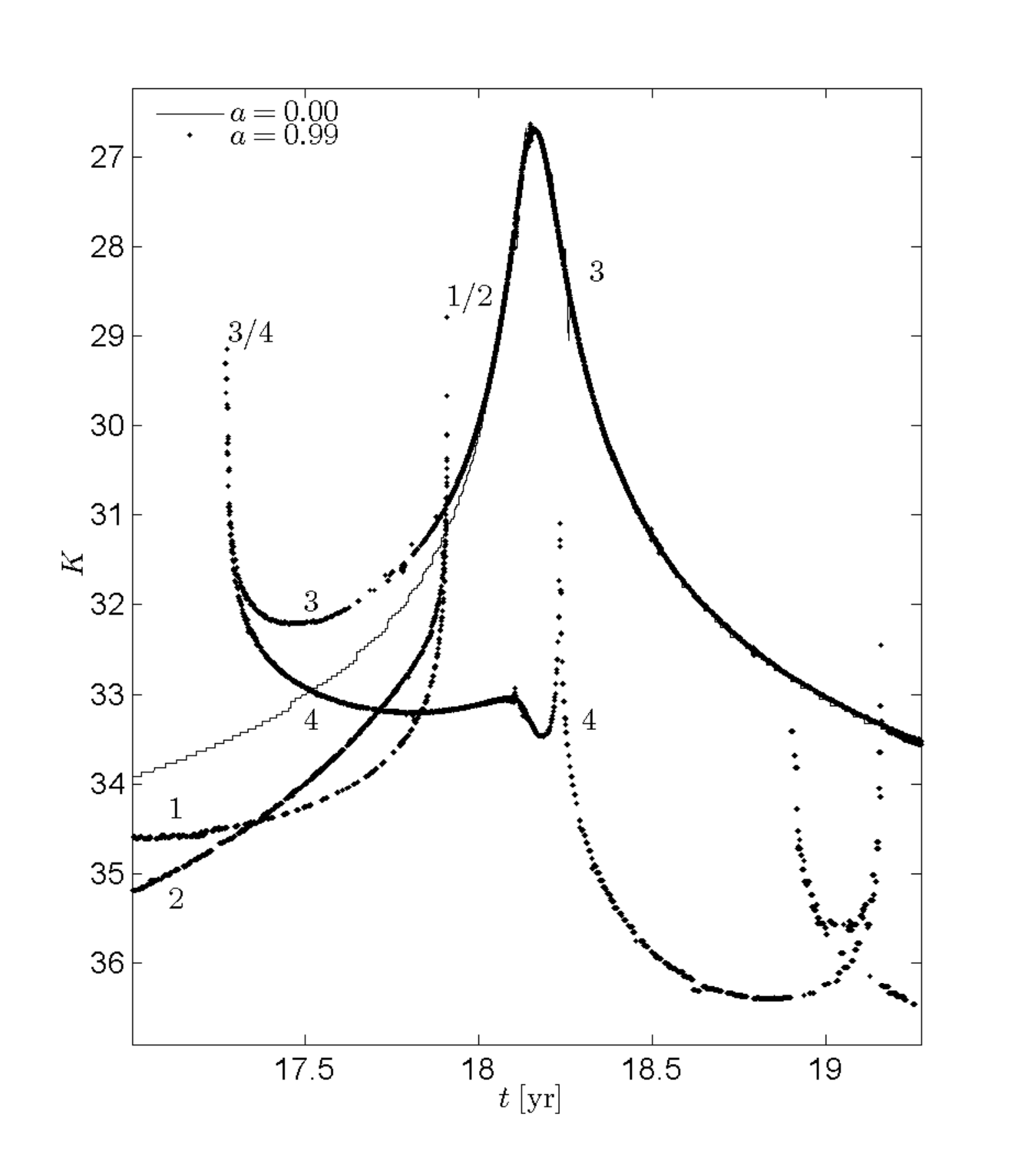}
\caption{The magnification of the secondary images of S2. The thin solid line shows the magnification of the secondary image of a non-rotating black hole. The dots represents the images when $a=0.99$. The numbers next to the lines corresponds to the numbers in Fig~\ref{fig:movement}: The spikes (1/2) and (3/4) correspond to the images at the point (1/2) and(3/4) in Fig~\ref{fig:movement}. The spike at (3/4) should be understood as an image that suddenly appears and then splits up into two separate images (3) and (4). While at the spike (1/2) the images (1) and (2) meets and then disappears.
The two spikes at 2019 are spikes in the magnitude of the tertiary image.}
\label{fig:magnitude}
\end{figure}

 
Our numerical work simulates the effects of a nonzero black hole angular momentum on the secondary images. We look for two such effects, namely its effect on the position of the secondary image and its effect on the magnification of the secondary image. The behaviour of the secondary (and higher order) images is, in general, very complicated, so as a first step we investigate the behaviour of the secondary image when the angular momentum is perpendicular to the galactic plane and $a=0.99$. Fig.~\ref{fig:movement} shows the position of the secondary (and tertiary) image at four different epochs in 2017, and Fig.~\ref{fig:magnitude} shows the magnification of these images as a function of time, when the angular momentum is perpendicular to the galactic plane.

Initially in Fig.~\ref{fig:movement}.A a secondary and tertiary image are visible at point 1 and 2, respectively. In Fig.~\ref{fig:movement}.B an image appears at point (3/4). This image corresponds to the spike (3/4) at 2017.3 in Fig.~\ref{fig:magnitude}. Subsequently, this image splits up into two images, resulting in a total of four images in Fig.~\ref{fig:movement}.C. These four images are visible as four separate lines at the 2017.5 epoch in Fig.~\ref{fig:magnitude}. The two original images (1) and (2) then merge into a single image in Fig.~\ref{fig:movement}.D and a spike in Fig.~\ref{fig:magnitude} and then disappears. The new images (3) and (4) subsequently become the new secondary and tertiary image and with time becomes image (1) and (2) in Fig.~\ref{fig:movement}.A. Also visible in Fig.~\ref{fig:magnitude} is a large peak in the secondary image in 2018 and two smaller spikes in the tertiary image in 2018-2019. These spikes will be important when we test the dependence of the spin axis.

The most interesting signature is the appearance of the two spikes in 2017, because they outshine possible secondary images of a non-rotating black hole and reach a magnitude of 29. Another interesting effect is the position of the secondary image seen in Fig.~\ref{fig:movement}, which turns away from the path of image in the non-rotating case. 
We will test how these effects depends on the angular momentum parameter $a$ and the orientation of the angular momentum vector.

\subsection{The spikes dependency on the the angular momentum}

\begin{figure}
\includegraphics[width=0.9\linewidth]{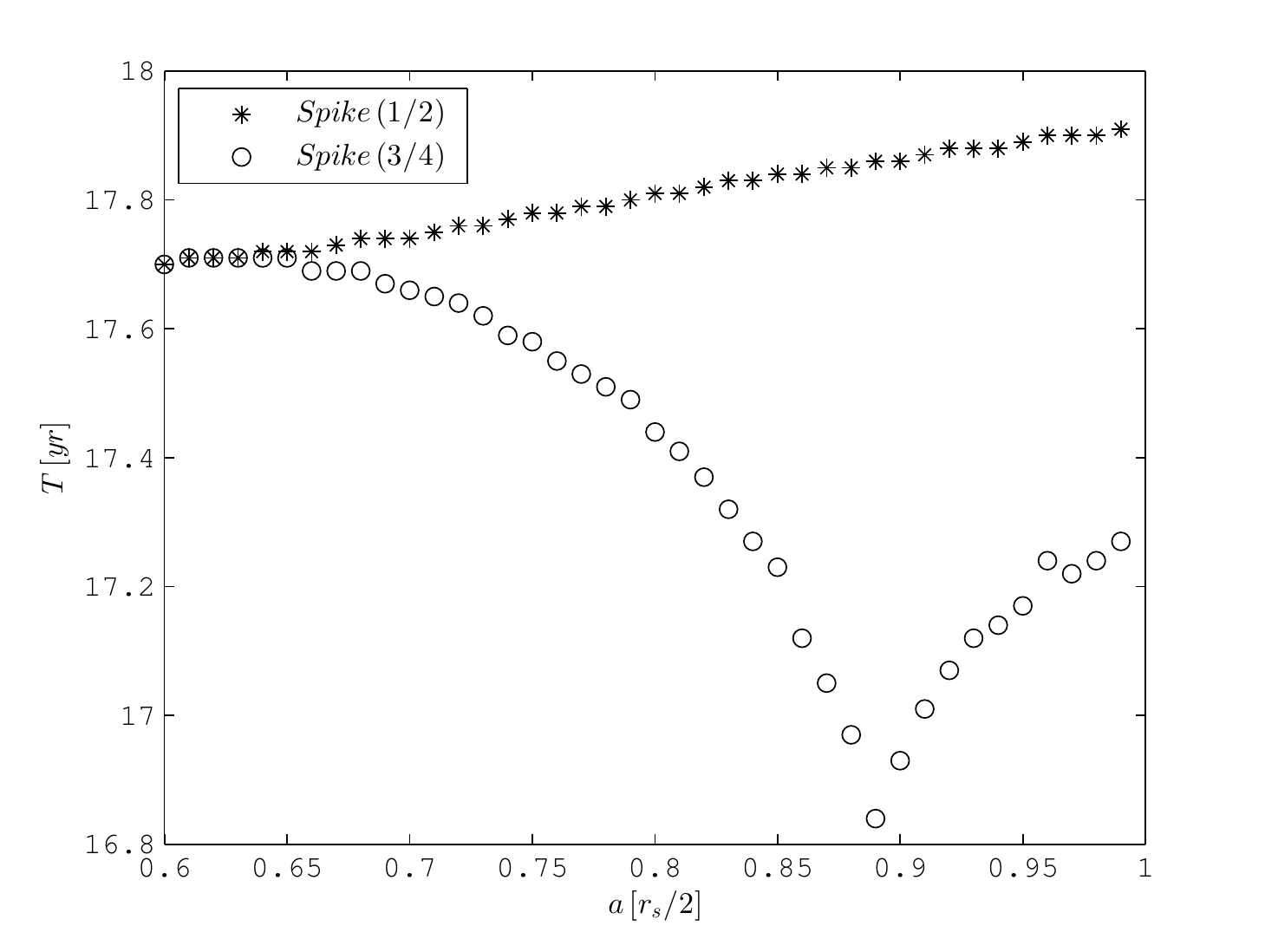}
\caption{The time of the two peaks in the secondary image as a function of the angular momentum paramater $a$ of the SMBH. The angular momentum vector is in this case perpendicular to the galactic plane, and rotating in the same direction as the galaxy.}
\label{fig:peaks}
\end{figure}

The timing of the appearance of the two spikes can be seen as a function of $a$ in Fig~\ref{fig:peaks}. The spikes are very faint and typically have a magnitude below $30$ in the K-band as long as they appear. However, when the spin is close to maximal the magnitude can reach $m \sim 29$, as demonstrated in Fig.~\ref{fig:magnitude}. 

It is possible to identify three qualitatively different cases for the spike structure:

\begin{itemize}
\item Below $a=0.63$ there are no spikes, just a peak of the tertiary image. 

\item In the region $0.63<a<0.90$ there are two spikes both belonging to the tertiary image. 

\item 
When $a>0.90$ the secondary and tertiary images mix at the two spikes, as demonstrated in Fig~\ref{fig:magnitude}. 
The mixing of the secondary and tertiary images happens when the time interval between the two spikes starts to decrease (see Fig~\ref{fig:peaks}). 
\end{itemize}

The spikes have proven to be robust to a variety of changes in the spin direction. We finally note that the peak magnitudes of the two spikes are hard to determine precisely because of the steep gradient around the appearance of the spike.

\subsection{Change of position}
Another effect is the change of position of the secondary image in Fig~\ref{fig:movement}. There is a clear change of the path of the secondary image when it bends inwards. To determine the duration of this shift in position we define an angle between the secondary image and compare to the secondary image for a non-rotating black hole. We then define the duration of the shift, $T_{\rm gap}$, as the time interval where the difference between the positions of the two images is larger than $45\, \text{degrees}$.

\begin{figure}
\includegraphics[width=0.9\linewidth]{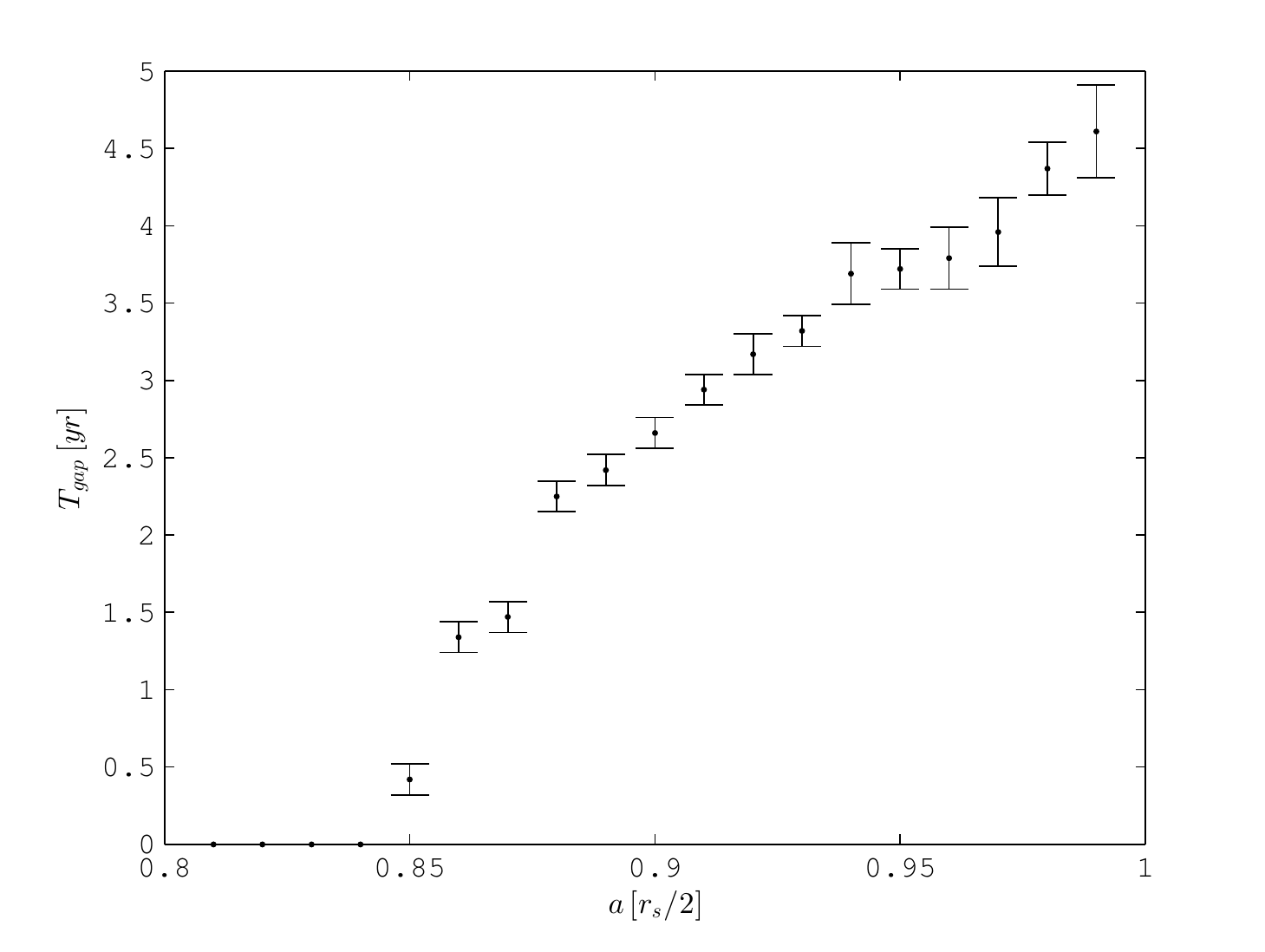}
\caption{Here the time the secondary position differs from that of a non-rotating black hole is shown as a function of the angular momentum parameter $a$. The spin direction is perpendicular to the galactic plane and rotates the same way as the galaxy.}
\label{fig:a_var}
\end{figure}

For the case of $a=0.99$ and the angular momentum perpendicular to the galactic plane $T_{\rm gap}=4.7\, years$. In Fig~\ref{fig:a_var} $T_{gap}$ can be seen as a function of $a$. It can be seen that the time where the image is shifted increases with $a$ and below $0.85$ the image is never shifted with more than $45$ degrees. 

The characteristic movement of the secondary image in Fig~\ref{fig:movement} where the image moves away from the circular movement of the $a=0$ case happens when $a>0.90$. This is consistent with the threshold value for when the secondary and tertiary image mix at the two spikes (1/2) and (3/4) in Fig~\ref{fig:magnitude}. And as shown in Fig~\ref{fig:peaks} the value for when the time between the two peaks starts to decrease. So the movement of the secondary and tertiary image starts to deviate from the $a=0$ case when $a>0.90$ and the two images starts to mix. For $0.85<a<0.90$ the image is shifted but remains on the same track. It is also shifted for $a<0.85$ but that shift is below the chosen limit of $45$ degrees.

\begin{figure*}
\includegraphics[width=0.95\linewidth]{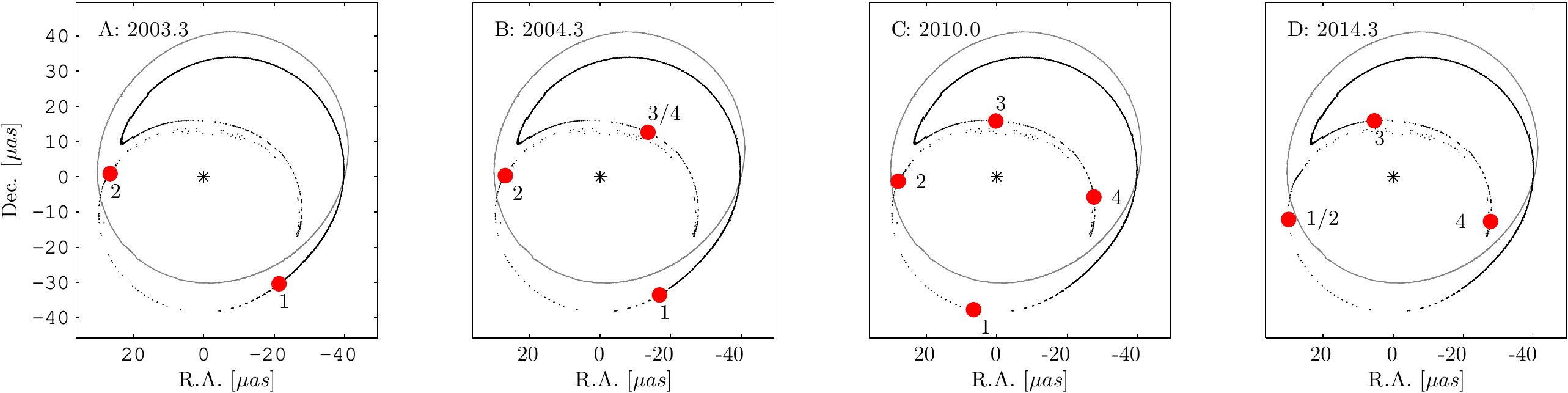}
\caption{Here the movement of the secondary image is shown for a rotation axis pointing at $\phi=1.50$ and $\theta_A=1.50$. On all four images, the position of the secondary image over the entire period is shown with a a black hole with $a=0$ and $a=0.99$. $A:$ in the year 2003.3 there is a secondary image at (1) and a tertiary image at (2). $B:$ in the year 2004.3 there is a secondary image at (1), a tertiary image at (2) and a image at (3/4) appear with high luminosity. $C:$ in the year 2010.0 there is a secondary image at (1) and a tertiary image at (2) the image at (3/4) splits into two images at (3) and (4). $D:$ in 2014.3 the two images (1) and (2) merges into a spike with high luminosity and disappear thereafter. Image (3) is now the secondary image and moves through the period and become image (1) in Fig.~A. Image (4) is the new tertiary image and ends up as image (2) in Fig.~A.}
\label{fig:gap2}
\end{figure*}

\subsection*{Dependence of spin axis}

To study the angular dependence of the gap more closely 
we have fixed the angular momentum parameter to $a=0.99 \, r_s /2$. The orientation of the spin axis was varied with angles $\phi_A$ and $\theta_A$ over the entire parameter space, ($0\leq \phi_A \leq 2 \pi$) and ($0\leq \phi_A \leq  \pi$), see Fig~\ref{fig:skema}. Where $\phi_A=0$ and $\theta_A=0$ corresponds to a rotation axis perpendicular to the galactic plane and rotating with the the galaxy, and $\phi_A=0$ and $\theta_A= \pi/2$ corresponds to the angular momentum vector pointing towards the solar system.

There is a large region around $\theta_A=0$ and $\phi_A=0$ where there is a gap like the one in Fig~\ref{fig:movement}, but there is another region around $\theta_A=\pi/2$ and $\phi_A=2.0$ where the gap is positioned as in Fig~\ref{fig:gap2}. The images in Fig~\ref{fig:gap2} also represent spikes in the points (1/2) and (3/4). These two spikes are related to the two small spikes around year 2019 in Fig~\ref{fig:magnitude}. When the spin axis is changed the secondary image starts to interact with these two spikes instead of the two spikes in year 2017.

\begin{figure}
\includegraphics[width=1\linewidth]{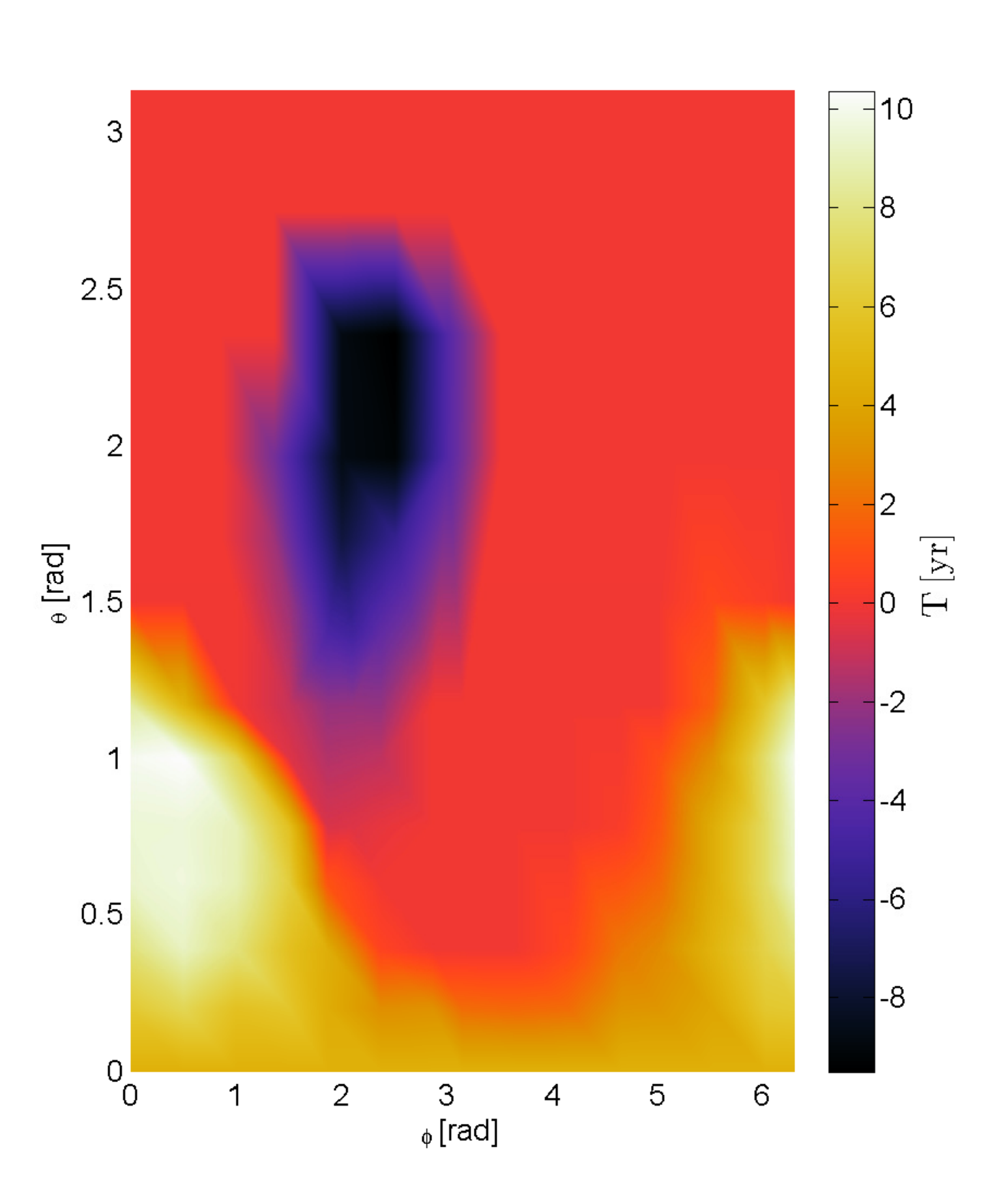}
\caption{Here is shown the length of the gaps dependent of the spin direction. If it is positive (red color scale) the gap is in the place of Fig~\ref{fig:movement}. If it is negative the gap is in the position of Fig~\ref{fig:gap2}.}
\label{fig:angles}
\end{figure}

In Fig~\ref{fig:angles} the duration of the gaps is shown for all spin orientations. For positive values the gap is positioned as in Fig~\ref{fig:movement} and for negative values it is positioned as in Fig~\ref{fig:gap2}.  
The magnitude of the secondary image is typically around $m \sim 32-35$, except at the spikes, and thus requires extreme sensitivity to observe.
The maximal gap size is at $\phi_A=1.00$ and $\theta_A= 0.39$, where the image is relocated for over $10$ years. 

\section{Observations}
Observations of the S-stars is usually done in the K-band because of the large amounts of dust. To complicate the observations of secondary images the Sgr A* has a diffuse source of radiation (spanning the K-band).
This, combined with the faint magnitude of the secondary images, puts high demands on both sensitivity and angular resolution of telescopes in order to observe these images.
Bozza and Mancini \citep{Bozza:2005sc} discuss the requirements to observe the tertiary image of S14 with a magnitude of $K=32.1$, and positioned inside the diffuse region. They conclude that the resolution should be below $0.1 \, \mu as$. To observe the spikes a similar resolution will be required.

Even if the required angular resolution can be reached the telescopes also need a very high sensitivity. Of current and planned telescopes GRAVITY and The Event Horizon Telescope (EHT) have the highest resolution. GRAVITY can observe in the K-band and can reach an angular resolution of $\sim 10\, \mu as$ \citep{Eisenhauer:2011}. 

EHT observe at sub-millimetre wavelengths and currently has an angular resolution of  $15 \,\mu as$. 
At sub-millimetre wavelength it is, however, not possible to reach the angular resolution of $\sim 0.1 \, \mu as$ with earth based telescopes because the baseline is limited to the diameter of the Earth. 
If observations of these spikes become possible, spectroscopy of the secondary image will still be necessary for confirmation.

The exact times for the appearance the spikes from S2 depend on the angular momentum direction of the SMBH. $\theta_A=0$ the spikes will occur in 2017  Fig~\ref{fig:peaks}. After that they will occur every 16th year as the star orbits the SMBH.

\section{Discussion and conclusion}
The secondary images of S2 and other S stars directly probe the space-time close to Sgr A*, which is necessary to investigate the spin of the SMBH.

Both the location and the magnitude of the secondary image of the star S2 depends on the spin and spin axis of the SMBH in SgrA*. The spikes can be used to put non-trivial constraints on the angular momentum and spin direction of the SMBH. 

To further constrain the parameters of the SMBH other S-stars could be used. They would likely be able to test other regions of the parameter space of Fig~\ref{fig:angles}.
The variation in the trajectories of the other stars mean that some will have spikes with a brighter magnitude, thus being easier to observe. We expect that the optimal trajectories have a stellar plane that has a small angle with the line sight to the solar system and optimally has the pericenter on the same side of the black hole as the sun.

However, we stress that observing the spikes for any of these stars will require a huge step forward in technology. 

In conclusion, we have demonstrated how secondary images from stars in orbit of the SMBH in the Milky Way can be used to examine the spin of the SMBH. Both the spin parameter $a$ and the spin axis of the SMBH in Sgr A* can influence the secondary images. This paper focuses on the secondary images of the star S2 as a worked example and we have demonstrated the appearance of spikes in the secondary image when the SMBH rotates. The spikes have a magnitude of $\sim29$ in the K-band. When the SMBH rotates with the rest of the galaxy the spikes appear when the spin parameter is large ($a>0.63$). The spikes are robust to changes to the spin direction.
Another interesting effect is the change of the secondary images position. 

Both features provide a very demanding observational challenge, but they also provide a very direct probe of the space time structure close to the central black hole of the Milky Way.

\bibliographystyle{mn2e}
\bibliography{paper}

\begin{thebibliography}{}

\bibitem[\protect\citeauthoryear{Ang\'elil \& Saha}{Ang\'elil \&
  Saha}{2014}]{Angelil:2014yea}
Ang\'elil R.,  Saha P.,  2014, MNRAS, 444, 3780

\bibitem[\protect\citeauthoryear{Boehle, Sch{\"o}del, Meyer \& Ghez}{Boehle
  et~al.}{2014}]{2014IAUS..303..242B}
Boehle A.,  Sch{\"o}del R.,  Meyer L.,    Ghez A.~M.,  2014, IAU Symposium,
  303, 242

\bibitem[\protect\citeauthoryear{Bozza \& Mancini}{Bozza \&
  Mancini}{2005}]{Bozza:2005sc}
Bozza V.,  Mancini L.,  2005, Astrophys.J., 627, 790

\bibitem[\protect\citeauthoryear{Brink, Geyer \& Hinderer}{Brink
  et~al.}{2015}]{Brink:2015roa}
Brink J.,  Geyer M.,    Hinderer T.,  2015, Phys.Rev., D91, 083001

\bibitem[\protect\citeauthoryear{Broderick, Fish, Doeleman \& Loeb}{Broderick
  et~al.}{2009}]{Broderick:2008sp}
Broderick A.~E.,  Fish V.~L.,  Doeleman S.~S.,    Loeb A.,  2009, Astrophys.
  J., 697, 45

\bibitem[\protect\citeauthoryear{Broderick, Fish, Doeleman \& Loeb}{Broderick
  et~al.}{2011}]{Broderick:2010kx}
Broderick A.~E.,  Fish V.~L.,  Doeleman S.~S.,    Loeb A.,  2011, Astrophys.
  J., 735, 110

\bibitem[\protect\citeauthoryear{Christian \& Loeb}{Christian \&
  Loeb}{2015}]{Christian:2015iia}
Christian P.,  Loeb A.,  2015, Phys.Rev., D91, 101301

\bibitem[\protect\citeauthoryear{Eisenhauer, Perrin, Brandner, Perraut, Amorin
  et~al.,}{Eisenhauer et~al.}{2011}]{Eisenhauer:2011}
Eisenhauer F.,  Perrin G.,  Brandner W.,  Perraut K.,  Amorin A.,    et~al.,
  2011, The Messenger, 143, 16

\bibitem[\protect\citeauthoryear{Ghez, Salim, Weinberg, Lu, Do et~al.,}{Ghez
  et~al.}{2008}]{Ghez:2008ms}
Ghez A.,  Salim S.,  Weinberg N.,  Lu J.,  Do T.,    et~al., 2008,
  Astrophys.J., 689, 1044

\bibitem[\protect\citeauthoryear{Gillessen, Eisenhauer, Trippe, Alexander,
  Genzel et~al.,}{Gillessen et~al.}{2009}]{Gillessen:2008qv}
Gillessen S.,  Eisenhauer F.,  Trippe S.,  Alexander T.,  Genzel R.,    et~al.,
  2009, Astrophys.J., 692, 1075

\bibitem[\protect\citeauthoryear{Jiang, Bambi \& Steiner}{Jiang
  et~al.}{2015}]{Jiang:2015dla}
Jiang J.,  Bambi C.,    Steiner J.~F.,  2015, preprint (arXiv:1504.01970)

\bibitem[\protect\citeauthoryear{Li \& Bambi}{Li \& Bambi}{2014}]{Li:2013jra}
Li Z.,  Bambi C.,  2014, JCAP, 1401, 041

\bibitem[\protect\citeauthoryear{Maoz}{Maoz}{1998}]{Maoz:1997yd}
Maoz E.,  1998, Astrophys.J., 494, L181

\bibitem[\protect\citeauthoryear{Melia, Bromley, Liu, Christopher \&
  Walker}{Melia et~al.}{2001}]{Melia:2001fp}
Melia F.,  Bromley B.~C.,  Liu S.,  Christopher   Walker K.,  2001,
  Astrophys.J., 554, L37

\bibitem[\protect\citeauthoryear{Meyer, Ghez, Schodel, Yelda, Boehle
  et~al.,}{Meyer et~al.}{2012}]{Meyer:2012hn}
Meyer L.,  Ghez A.,  Schodel R.,  Yelda S.,  Boehle A.,    et~al., 2012,
  Science, 338, 84

\bibitem[\protect\citeauthoryear{Moderski \& Sikora}{Moderski \&
  Sikora}{1996}]{1996MNRAS.283..854M}
Moderski R.,  Sikora M.,  1996, MNRAS, 283, 854

\bibitem[\protect\citeauthoryear{Muller \& Grave}{Muller \&
  Grave}{2009}]{Muller:2009bw}
Muller T.,  Grave F.,  2009, preprint (arXiv:0904.4184)

\bibitem[\protect\citeauthoryear{Penrose}{Penrose}{1969}]{Penrose:1969pc}
Penrose R.,  1969, Riv.Nuovo Cim., 1, 252

\bibitem[\protect\citeauthoryear{Reynolds}{Reynolds}{2013}]{Reynolds:2013rva}
Reynolds C.~S.,  2013, Class.Quant.Grav., 30, 244004

\bibitem[\protect\citeauthoryear{Volonteri, Madau, Quataert \& Rees}{Volonteri
  et~al.}{2005}]{Volonteri:2004cf}
Volonteri M.,  Madau P.,  Quataert E.,    Rees M.~J.,  2005, Astrophys.J., 620,
  69

\bibitem[\protect\citeauthoryear{Wald}{Wald}{1984}]{Wald:1984}
Wald R.,  1984, {General Relativity}.
University Of Chicago Press

\bibitem[\protect\citeauthoryear{Zuo, Wu, Fan, Green, Wang et~al.,}{Zuo
  et~al.}{2015}]{Zuo:2014fea}
Zuo W.,  Wu X.-B.,  Fan X.,  Green R.,  Wang R.,    et~al., 2015, Astrophys.J.,
  799, 189

\end{thebibliography}

\end{document}